# DIGITAL IMAGING AND ANALYSIS OF DUSTY PLASMAS

C. M. Boessé, M. K. Henry, T. W. Hyde, and L. S. Matthews

*CASPER (Center for Astrophysics, Space Physics, and Engineering Research)*
*Baylor University, P.O. Box 97310, Waco, TX 76798-7310, USA*

## ABSTRACT

Dust particles immersed within a plasma environment, such as those found in planetary rings or cometary environments, will acquire an electric charge. If the ratio of interparticle potential energy to average kinetic energy is high enough the particles will form either a 'liquid' structure with short-range ordering or a crystalline structure with long-range ordering. Since their discovery in laboratory environments in 1994, such crystals have been the subject of a variety of experimental, theoretical, and numerical investigations. Laboratory experiments analyzing the behavior of dust grains in a plasma rely on optical diagnostics to provide data about the system in a non-perturbative manner. In the past, capturing, imaging, and analyzing crystalline structure in dusty plasmas has been a non-trivial problem. Utilizing digital imaging and analysis systems, data capture, image formatting, and analysis can be done quickly. Following data capture, image analysis is conducted using modified Particle Image Velocimetry (PIV) and Particle Tracking Velocimetry (PTV) algorithms. The data extracted is then used to construct Voronoi diagrams, calculate particle density, inter-particle spacing, pair correlation functions, and thermal energy. From this data other dust plasma parameters can be inferred such as inter-particle forces and grain charges.

## INTRODUCTION

Studies of complex dusty plasmas and associated phenomena are known to have applications in many astrophysical environments, such as protoplanetary systems, cometary tails, and planetary rings, as well as in laboratory environments, primarily those which arise in the plasma-aided etching of silicon wafers (Thomas, 1994). In laboratory environments, dusty plasmas form when the plasma temperature is low and dust is introduced into the plasma. The immersed dust then generally acquires a high negative charge from plasma fluxes to the dust grain surfaces. The charged dust particles interact via shielded Coulomb potentials and may form a regular ordered array if grain size and charge dispersion are small (Pieper, 1996a). The recent discovery of this Coulomb crystallization has led to advances in the understanding of dynamics and phase transitions in solids, as well as a better understanding of the interactions between the strongly coupled dust grains (Pieper, 1996a; Thomas, 1994). This has resulted in a rapidly growing body of literature published on the topic (Barkan, 1994; Melzer, 2000; Pieper, 1996a).

Both experimental and theoretical dusty plasma research investigate the charging process of dust grains within dusty plasmas and the resulting dynamics of the complex system. Of these, the formation of lattice structure and the perturbation of grains within the crystal are two of the most interesting (Barkan, 1994). In laboratory settings, different lattice structures in dust crystals, including close packed hexagonal, body centered cubic, and face centered cubic have been observed (Pieper, 1996a; Pieper, 1996b; Morfill, 1999; Quinn, 1996). Perturbation effects such as Mach cones and wave propagation are also becoming subjects of interest experimentally (Melzer, 2000; Homann, 1998; Samsonov, 1999; Samsonov, 2000; Smith, 2002; Zuzic, 1996). Dust particles have been found to form ordered structures under various types of conditions, and several parameters have been established to describe these experimentally (Thomas, 1994; Pieper, 1996a; Pieper, 1996b). Two of these are $\Gamma$, the ratio of the average potential energy of the dust grains to the thermal energy, and $\kappa$, the ratio of the interparticle spacing to the Debye length, $\lambda$, which helps define the effective shielding radius around a charged dust grain. Efforts have recently been made to associate the properties of dust crystals to the values of these parameters (Pieper, 1996b).

In order to create a controlled plasma environment and make meaningful quantitative and qualitative observations of these systems, the GEC rf reference cell was introduced as a standard tool in which to produce and collect data (Hargis, et al., 1994; Samsonov, 2000). Obtaining and analyzing data from the reference cell without perturbing the plasma requires an efficient optical diagnostic system. In the process outlined in this paper, images are taken straight to digital format by means of a frame grab board with MATLAB scientific computing software then used to analyze the resulting data. Particle Image Velocimetry (PIV) algorithms have been used to analyze the motion of dust grains in a plasma (Thomas, 2001) while other studies tracked individual particles via Particle Tracking Velocimetry (PTV) in order to determine the direction and magnitude of the particle flow (Samsonov, 2000). The system described here utilizes both methods and then integrates the data giving estimated velocity vectors and/or statistical estimations of intensity shifts.

**METHODS**

**GEC Reference Cell**

Experimental procedures for the formation of dust crystals are conducted in a GEC reference cell utilizing an argon rf discharge of 13.56 MHz (Hargis, 1994). The discharge is created using two strongly biased electrodes, placed two centimeters apart, in the plasma chamber. The lower electrode is a disk 10 cm in diameter, while the upper electrode is a ring of 10 cm outer diameter and 9 cm inner diameter. Once the plasma is established, melamine formaldehyde particles 8.9 μm in diameter are introduced into the plasma environment. The discharge region is illuminated by vertical and horizontal diode laser sheets, and imaged using two CCD cameras--one mounted horizontally and another mounted vertically above the cell, as can be seen in Figure 1. Both cameras utilize band pass filters in order to view particles within the illuminated planes while minimizing ambient light due to glow discharge and are focused on the laser sheet perpendicular to their respective observation positions.

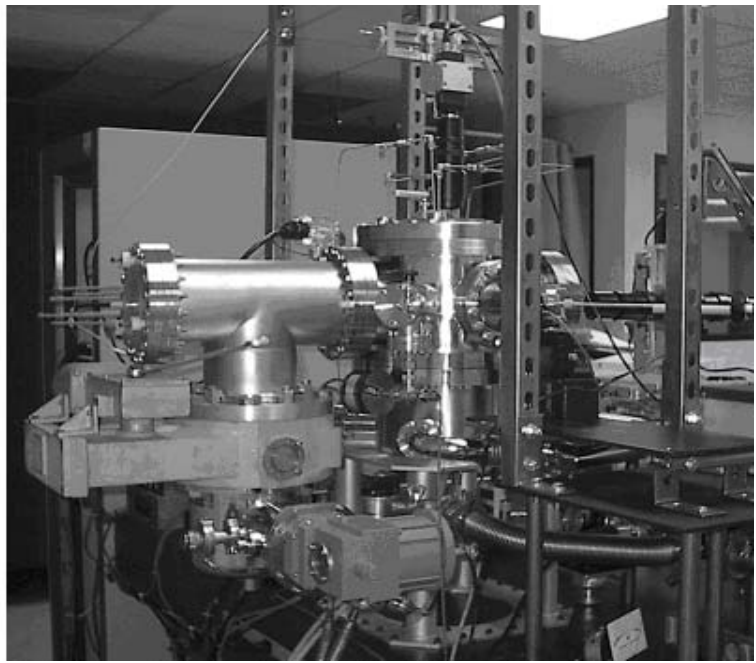

Fig. 1. GEC reference cell and optical system of two cameras mounted at right angles – one top-mounted and one side-mounted.

For the upper camera, a large field of view is desirable in order to best observe long-range crystalline structure. Crystals formed from dust particles of seven to ten microns have been observed having interparticle spacing of 128-250 micrometers (Pieper, 1996a; Thomas, 1994). To adequately resolve individual particles within such a crystal lattice, the top mounted camera utilizes a Close Focus brand zoom lens with a variable focal length of 18-108 mm, a six to one magnification ratio, and a lens magnification doubler which produces a field of view of approximately 11 mm by 11mm.

The horizontal lattice structure is observed from the upper camera, while the separation distance between vertical layers is observed from the side camera. As vertical layer spacing, approximately 20-100 μm (Pieper, 1996), is much smaller than horizontal interparticle spacing, approximately 128-250 μm, an Infinity brand K2 long distance microscope lens system was utilized on the side-mounted camera, which allowed for more detailed observations of particle positions and perturbative effects. The primary use of the side mounted microscopic lens is to determine if a crystal consists of individual particles or conglomerates and the results of external perturbations on individual particles.

**Digital Imaging and Analysis**

Data images are captured and transferred by means of a Coreco PC-RGB frame grabber board. The frame grabber board has the capability of transferring images of 1600 by 1600 pixels from each camera at the rate of 30 frames per second, interlaced. Each image from the camera contains 512 by 480 pixels and occupies approximately 250 Kb of disk space as bitmap files. Image data consists of a directory of grayscale intensity images with pixel values ranging from zero to one (Figure 2). Prior to image capture and transfer, the contrast may be set such that there is maximum differentiation between particles and background. This reduces the effects of digital filtering during analysis, which may lead to loss of data.

The image data were analyzed using an image segmentation algorithm, as well as both PIV (Particle Image Velocimetry) and PTV (Particle Tracking Velocimetry) techniques. Generally, PIV algorithms are used to obtain data from a densely seeded field, where individual particles cannot be easily distinguished. PTV techniques are used when the field is sparsely seeded and particles are well defined. The limiting factor for the utility of PTV is interparticle spacing since when the dust grains are closely situated PTV algorithms are inefficient. When particle velocities are such that they move over half the interparticle spacing, the probability of positively identifying such a particle in sequential images is unreliable. Essentially, when all particles in an image remain within their respective voronoi cells in sequential images, PTV techniques can be used reliably.

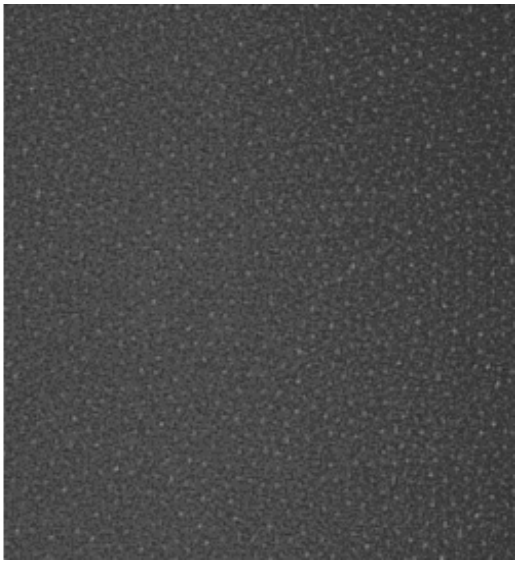
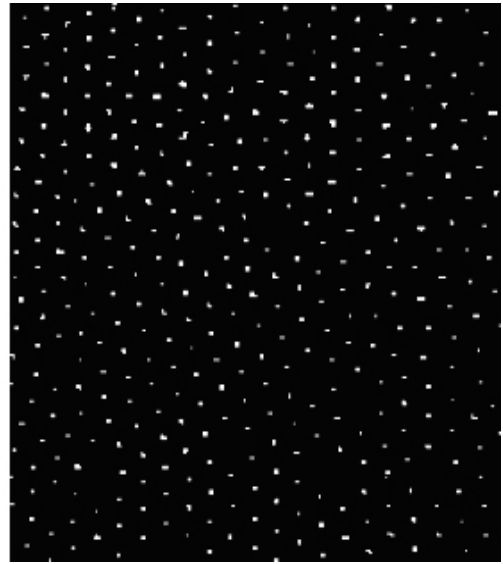

Fig. 2. Raw unfiltered digital image of dusty plasma

Fig. 3. Grayscale image converted to black and white and magnified to show ordering.

In the first data analysis technique, the images are analyzed using an image segmentation process where image contrast is maximized (Liberzon, 2001). The image is then converted to a binary form using a threshold intensity where all values below the threshold value are set to zero, resulting in a black pixel, while all pixel values above the threshold value are set to one, resulting in a white pixel as seen in Figure 3. In order to identify individual dust grains, groups of adjacent white pixels are identified and then determined to be either valid dust grains or noise by means of a minimum pixel filter. Adjacent white pixels which meet or exceed the filter size are considered dust grains, while other pixel groupings less than the filter size are ignored and are considered noise or representations of dust in other parallel planes. A list of the positions and sizes of the dust particles is then stored in memory. From this list of particle positions, interparticle spacing and pair correlation functions are determined (Fig. 4a) and a Voronoi diagram is constructed (Fig. 4b).

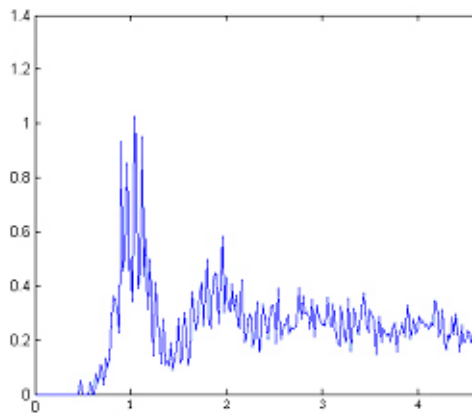
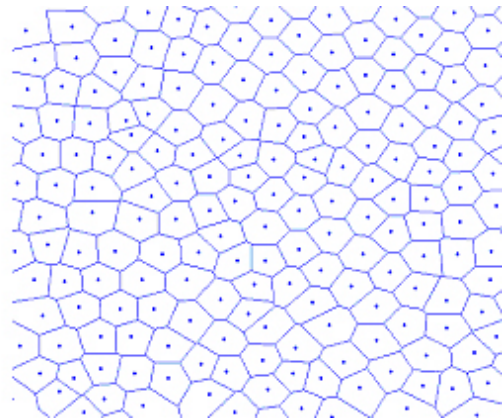

Fig. 4a.  Pair correlation function with respect to interparticle distance

Fig. 4b.  Voronoi diagram

After the particles have been located in each image, the list of particle positions for each pair of sequential images is compared in order to detect the motion of each dust particle. This particle tracking velocimetry (PTV) technique implements a radar-type detection system to find where each particular particle is represented in the next image. Knowing the location of a particular particle in the first image, sequentially larger areas centered about that same point could be examined in the second image for the particle. The search area begins with a radius of ten percent of the interparticle distance and grows by increments of ten percent until it finds a particle. If no particle is found after the search area radius has grown to fifty percent of the interparticle distance, the particle is considered lost. If two particles are found in a single search area, the closest one is chosen. When a particle is found, the system verifies by the same technique that the particle in the first image is the closest particle in that image to the particle found in the second image. Using the PTV analysis technique on data of known wave position and velocity as a test of the routine, the algorithm produced the required values (Goree, 1996). By determining the distance and direction of the motion of each particle, along with the known time step between frames, the velocity information for each particle can be calculated and a PTV velocity vector plot generated, as shown in Figure 5. After confirmation of the analysis technique, the routine was applied to experimental data obtained from the CASPER GEC reference cell. Random velocities with a maximum amplitude of 2.0 mm/sec and no propagated waves were obtained.

The PIV technique, which is a separate algorithm, does not make use of the list of particle positions or the image filtering system, but rather divides the image into a user-specified number of equal sized squares, or interrogation regions. Each interrogation region is then assigned a velocity vector based on the collective shift of pixel intensity in that region from one frame to the next. In this way, average velocities for small groups of dust grains are calculated rather than individual particle velocities. Because the image is not converted to binary, more noise is present in the image, and a signal to noise limit, $l$, must be set. This means that when an intensity peak is found in an interrogation region, it must be at least $l$ times the intensity of the average value of the pixels in that region in order to be considered valid. Also, an upper limit for the velocity magnitude may be set based on the time step and image size. After these validation processes, a list of region velocities and a PIV velocity vector plot is generated, as shown in Figure 6. PIV has been shown to have advantages over many conventional techniques for selected experimental conditions (Thomas, 1999).

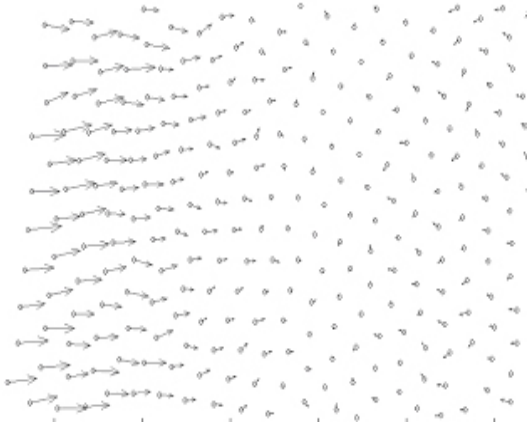 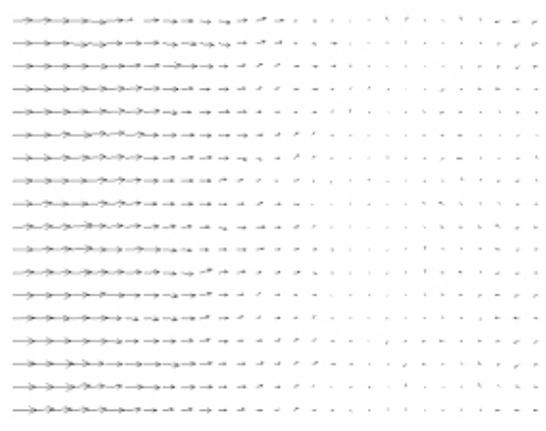

Fig. 5. PTV velocity vector plot	Fig. 6. PIV velocity vector plot

## RESULTS

Data analysis generated Voronoi diagrams and pair correlation functions, and two types of velocity vector plots, which prove useful for observing the crystalline structure formed by the dust grains. The pair correlation function seen in Figure 4a has two well-defined peaks, indicating a high degree of order at a distance of at least the second nearest neighbor in the horizontal plane. The Voronoi diagram (Figure 4b) displays 54 five-sided cells, 127 six-sided cells, and 30 seven-sided cells (a 1.5:4:1 ratio) where the high number of six-sided Wigner-Seitz cells indicates a highly ordered system of two dimensional hexagonal lattice points. The information provided by Voronoi diagrams and the pair correlation function demonstrates a large amount of order in the crystalline lattice. PIV and PTV analysis of the video of the dust crystal demonstrated a longitudinal wave propagating horizontally through the crystal. Position of this wave is represented in the velocity vector plots (Figures 5 and 6). The crest of this wave is on the right side of the image, represented in the plot as a vertical section of large horizontal velocity vectors. The left side of the images show the trough of the wave represented by velocity vectors of nearly zero magnitude. Graphical data provided by the Matlab program make it possible to describe the discrete velocities of the dust grains as well as the collective dynamics of the system.

## CONCLUSIONS

The development and implementation of a completely digital optical and analysis system, which minimizes the loss of data in all stages of acquisition and processing, has been completed. This allows data to be acquired much more quickly and efficiently than in previous methods. Analysis is also greatly improved, due primarily to uncompromised raw data. Matlab provides a user-friendly analysis tool for processing data via PIV and PTV algorithms. PIV algorithms have the advantage of greatly reducing computation time and the ability to calculate particle velocities when individual particles are not resolvable, while PTV algorithms provide more specific information about dust grain interactions and more precise velocity calculations. The current analysis program is unique in that it performs both PIV and PTV techniques and can be programmed to dynamically decide which algorithm is most appropriate for a certain image frame. In future developments, a hybrid adaptive scheme will be integrated into the process, which will utilize the PIV data as a validation technique for PTV data and data from the two cameras will be utilized to produce three dimensional diagrams of the crystal lattice and provide further information concerning interparticle forces and grain charging processes.